\documentclass[runningheads]{llncs}
\usepackage[T1]{fontenc}
\usepackage{graphicx}
\usepackage{amsfonts}
\usepackage{amsmath}
\usepackage[dvipsnames]{xcolor}
\usepackage{multirow}
\usepackage{array}
\usepackage{caption} 
\usepackage{eso-pic} 

\newcolumntype{C}[1]{>{\centering\arraybackslash}m{#1}}

\begin{document}

\AddToShipoutPictureBG*{%
  \AtPageUpperLeft{%
    \hspace{\paperwidth}%
    \put(6,-\paperwidth/20){%
      \parbox{\paperwidth}{%
        \raggedright
        \textcolor{red}{Accepted for publication in MICCAI 2023. \\
        Cite as: Yalcinkaya DM, et al. ``Temporal Uncertainty Localization to Enable Human-in-the-loop Analysis of Dynamic Contrast-enhanced Cardiac MRI Datasets''. \textit{Medical Image Computing and Computer-Assisted Intervention (MICCAI)}, pp. 453-462, 2023. In press. \\DOI: 10.1007/978-3-031-43898-1\_44}
}}}}%

\title{Temporal Uncertainty Localization to Enable Human-in-the-loop Analysis of Dynamic Contrast-enhanced Cardiac MRI Datasets}
\titlerunning{Temporal Uncertainty Localization for DCE-CMRI Datasets}
%
\author{Dilek M. Yalcinkaya\inst{1,2} \and
Khalid Youssef\inst{1,3} \and
Bobak Heydari\inst{4} \and
Orlando Simonetti\inst{5} \and
Rohan Dharmakumar\inst{3,6} \and
Subha Raman\inst{3,6} \and
Behzad Sharif\inst{1,3,6}}



\authorrunning{Yalcinkaya et al.}
%
\institute{Laboratory for Translational Imaging of Microcirculation, Indiana University School of Medicine (IUSM), Indianapolis, IN, USA \and
Elmore Family School of Electrical \& Computer Engineering, Purdue University, West Lafayette, IN, USA \and
Krannert Cardiovascular Research Center, IUSM/IU Health Cardiovascular
Institute, Indianapolis, IN, USA \and
Stephenson Cardiac Imaging Centre, University of Calgary, Alberta, Canada \and
Department of Internal Medicine, Division of Cardiovascular Medicine, Davis Heart and Lung Research Institute, The Ohio State University, Columbus, OH, USA \and
Weldon School of Biomedical Eng., Purdue University, West Lafayette, IN, USA}

\maketitle            
\begin{abstract}
Dynamic contrast-enhanced (DCE) cardiac magnetic resonance imaging (CMRI) is a widely used modality for diagnosing myocardial blood flow (perfusion) abnormalities. During a typical free-breathing DCE-CMRI scan, close to 300 time-resolved images of myocardial perfusion are acquired at various contrast ``wash in/out'' phases. Manual segmentation of myocardial contours in each time-frame of a DCE image series can be tedious and time-consuming, particularly when non-rigid motion correction has failed or is unavailable. While deep neural networks (DNNs) have shown promise for analyzing DCE-CMRI datasets, a ``dynamic quality control'' (dQC) technique for reliably detecting failed segmentations is lacking. Here we propose a new space-time uncertainty metric as a dQC tool for DNN-based segmentation of free-breathing DCE-CMRI datasets by validating the proposed metric on an external dataset and establishing a human-in-the-loop framework to improve the segmentation results. In the proposed approach, we referred the top 10\% most uncertain segmentations as detected by our dQC tool to the human expert for refinement. This approach resulted in a significant increase in the Dice score (p<$0.001$) and a notable decrease in the number of images with failed segmentation (16.2\% to 11.3\%) whereas the alternative approach of randomly selecting the same number of segmentations for human referral did not achieve any significant improvement. Our results suggest that the proposed dQC framework has the potential to accurately identify poor-quality segmentations and may enable efficient DNN-based analysis of DCE-CMRI in a human-in-the-loop pipeline for clinical interpretation and reporting of dynamic CMRI datasets.

\end{abstract}
\section{Introduction}
Dynamic contrast-enhanced (DCE) cardiac MRI (CMRI) is an established medical imaging modality for detecting coronary artery disease and stress-induced myocardial blood flow abnormalities. Free-breathing CMRI protocols are preferred over breath-hold exam protocols due to the greater patient comfort and applicability to a wider range of patient cohorts who may not be able to perform consecutive breath-holds during the exam. Once the CMRI data is acquired, a key initial step for accurate analysis of the DCE scan is contouring or segmentation of the left ventricular myocardium. In settings where non-rigid motion correction (MoCo) fails or is unavailable, this process can be a time-consuming and labor-intensive task since a typical DCE scan includes over 300 time frames.  

Deep neural network (DNN) models have been proposed as a solution to this exhausting task~\cite{scannell2020deep,chen2020improving,xue2020automated,youssef2021}. However, to ensure trustworthy and reliable results in a clinical setting, it is necessary to identify potential failures of these models. Incorporating a quality control (QC) tool in the DCE image segmentation pipeline is one approach to address such concerns. Moreover, QC tools have the potential to enable a human-in-the-loop framework for DNN-based analysis~\cite{mozannar2020consistent}, which is a topic of interest especially in medical imaging ~\cite{budd2021survey,ejom}. In a human-A.I collaboration framework, time/effort efficiency for the human expert should be a key concern. For free-breathing DCE-CMRI datasets, this time/effort involves QC of DNN-derived segmentations for each time frame. Recent work in the field of medical image analysis~\cite{wang2019aleatoric,wickstrom2020uncertainty,hoebel2020exploration,devries2018leveraging,mehrtash2020confidence,roy2019bayesian} and specifically in  CMRI~\cite{ieeetmi,puyol2020automated,hann2021deep,yalcinkaya2021deep,sander2020automatic} incorporate QC and uncertainty assessment to assess/interpret DNN-derived segmentations. Still, a QC metric that can both temporally and spatially localize uncertain segmentation is lacking for dynamic CMRI.
 
Our contributions in this work are two-fold: (i) we propose an innovative spatiotemporal dynamic quality control (dQC) tool for model-agnostic test-time assessment of DNN-derived segmentation of free-breathing DCE CMRI; (ii) we show the utility of the proposed dQC tool for improving the performance of DNN-based analysis of external CMRI datasets in a human-in-the-loop framework. Specifically, in a scenario where only 10\% of the dataset can be referred to the human expert for correction, although random selection of cases does not improve the performance (p=n.s. for Dice), our dQC-guided selection yields a significant improvement (p<0.001 for Dice). To the best of our knowledge, this work is the first to exploit the test-time agreement/disagreement between spatiotemporal patch-based segmentations to derive a dQC metric which, in turn, can be used for human-in-the-loop analysis of dynamic CMRI datasets. 

\section{Methods}

\subsection{Training/testing dynamic CMRI datasets} 
Our training/validation dataset (90\%/10\% split) consisted of DCE CMRI (stress first-pass perfusion) MoCo image-series from 120 subjects, which were acquired using 3T MRI scanners from two medical centers over 48-60 heartbeats in 3 short-axis myocardial slices~\cite{zhou2017first}. The training set was extensively augmented by simulating breathing motion patterns and artifacts in the MoCo image-series, using random rotations ($\pm50^\circ$), shear ($\pm10^\circ$), translations (±2 pixels), scaling (range: [0.9, 1.1]), flat-field correction with 50\% probability ($\sigma\in$ [0, 5]), and gamma correction with 50\% probability ($\gamma\in$ [0.5, 1.5]). To assess the generalization of our approach, an external dataset of free-breathing DCE images from 20 subjects acquired at a third medical center was used. Local Institutional Review Board approval and written consent were obtained from all subjects. 

\subsection{Patch-based quality control}
Patch-based approaches have been widely used in computer vision applications for image segmentation ~\cite{bai2013probabilistic,coupe2011patch} as well as in the training of deep learning models~\cite{kuo2019expert,fahmy2020three,rudie2021three,lu2021data,hou2016patch}. In this work, we train a spatiotemporal (2D+time) DNN to segment the myocardium in DCE-CMRI datasets. Given that each pixel is present in multiple patches, we propose to further utilize this patch-based approach at test-time by analyzing the discordance of DNN inference (segmentation output) of each pixel across multiple overlapping patches to obtain a dynamic quality control map.

Let $\mathrm{\Theta}(w)$ be a patch extraction operator decomposing dynamic DCE-CMRI image $I\in\mathbb{R}^{M\times N\times T}$ into spatiotemporal patches $\theta\in \mathbb{R}^{K \times K \times T}$ by using a sliding window with a stride $w$ in each spatial direction. Also, let $\mathrm{\Gamma}_{m,n}$ be the set of overlapping spatiotemporal patches that include the spatial location $(m,n)$ in them. Also, $p^i_{m,n}(t)\in\mathbb{R}^T$ denotes the segmentation DNN's output probability score for the $i^{th}$ patch at time $t$ and location $(m,n)$. The binary segmentation result $\mathcal{S}$ $\in\mathbb{R}^{M\times N\times T}$ is derived from the mean of the probability scores from the patches that are in $\mathrm{\Gamma}_{m,n}$ followed by a binarization operation. Specifically, for a given spatial coordinate $(m,n)$ and time t, the segmentation solution is:

\begin{equation}
    \mathcal{S}_{m,n}(t) = 
    \begin{cases}
    1, & \text{if $\frac{1}{|\mathrm{\Gamma}_{m,n}|}\sum_{i=1}^{|\mathrm{\Gamma}_{m,n}|} p^i_{m,n}(t)$ $\geq 0.5$}.\\
    0, & \text{otherwise}.
    \end{cases}
\end{equation}
The patch-combination operator, whereby probability scores from multiple overlapping patches are averaged, is denoted by $\mathrm{\Theta}^{-1}(w)$.

The dynamic quality control (dQC) map $\mathcal{M}$ $\in\mathbb{R}^{M\times N\times T}$ is a space-time object and measures the discrepancy between different segmentation solutions obtained at space-time location $(m,n,t)$ and is computed as:

\begin{equation}
\mathcal{M}_{m,n}(t) = \texttt{std}(p^1_{m,n}(t), p^2_{m,n}(t), \dots , p^{|\mathrm{\Gamma}_{m,n}|}_{m,n}(t))
\end{equation}
 
\noindent where \texttt{std} is the standard deviation operator. Note that to obtain $\mathcal{S}$ and $\mathcal{M}$, the same patch combination operator $\mathrm{\Theta}^{-1}(w)$ was used with $w_{\mathcal{M}}<w_{\mathcal{S}}$. Further, we define 3 quality-control metrics based on $\mathcal{M}$ that assess the segmentation quality at different spatial levels: pixel, frame, and slice (image series). First,  $\mathcal{Q}^\mathrm{pixel}_{m,n}(t)\in\mathbb{R}$ is the value of $\mathcal{M}$ at space-time location $(m,n,t)$ normalized by the segmentation area at time $t$:
\begin{equation}
\mathcal{Q}^\mathrm{pixel}_{m,n}(t) := \frac{\mathcal{M}_{m,n}(t)}{\sum_{m,n} \mathcal{S}_{m,n}(t)}
\tag{3.1}\label{eq:3.1}
\end{equation}

\noindent Next, $\mathcal{Q}^\mathrm{frame}(t)\in\mathbb{R}^T$ quantifies the \emph{per-frame segmentation uncertainty} as per-frame energy in $\mathcal{M}$ normalized by the corresponding per-frame segmentation area at time $t$:
\begin{equation}
\mathcal{Q}^\mathrm{frame}(t) := \frac{\|\mathcal{M}(t)\|_F}{\sum_{m,n} \mathcal{S}_{m,n}(t)}
\tag{3.2}\label{eq:3.2}
\end{equation}

\noindent where $\|\cdot\|_F$ is the Frobenius norm and $\mathcal{M}(t)\in\mathbb{R}^{M\times N}$ denotes frame $t$ of the dQC map $\mathcal{M}$. 
Lastly, $\mathcal{Q}^{\mathrm{slice}}$ assesses the overall segmentation quality of the acquired myocardial slice (image series) as the average of the per-frame metric along time:

\begin{equation}
\mathcal{Q}^\mathrm{slice} := \frac{1}{T}\sum_{t=1}^{T} \mathcal{Q}^\mathrm{frame}(t)
\tag{3.3}\label{eq:3.3}
\end{equation}

\begin{figure}
\centering
\includegraphics[width=4.5in]{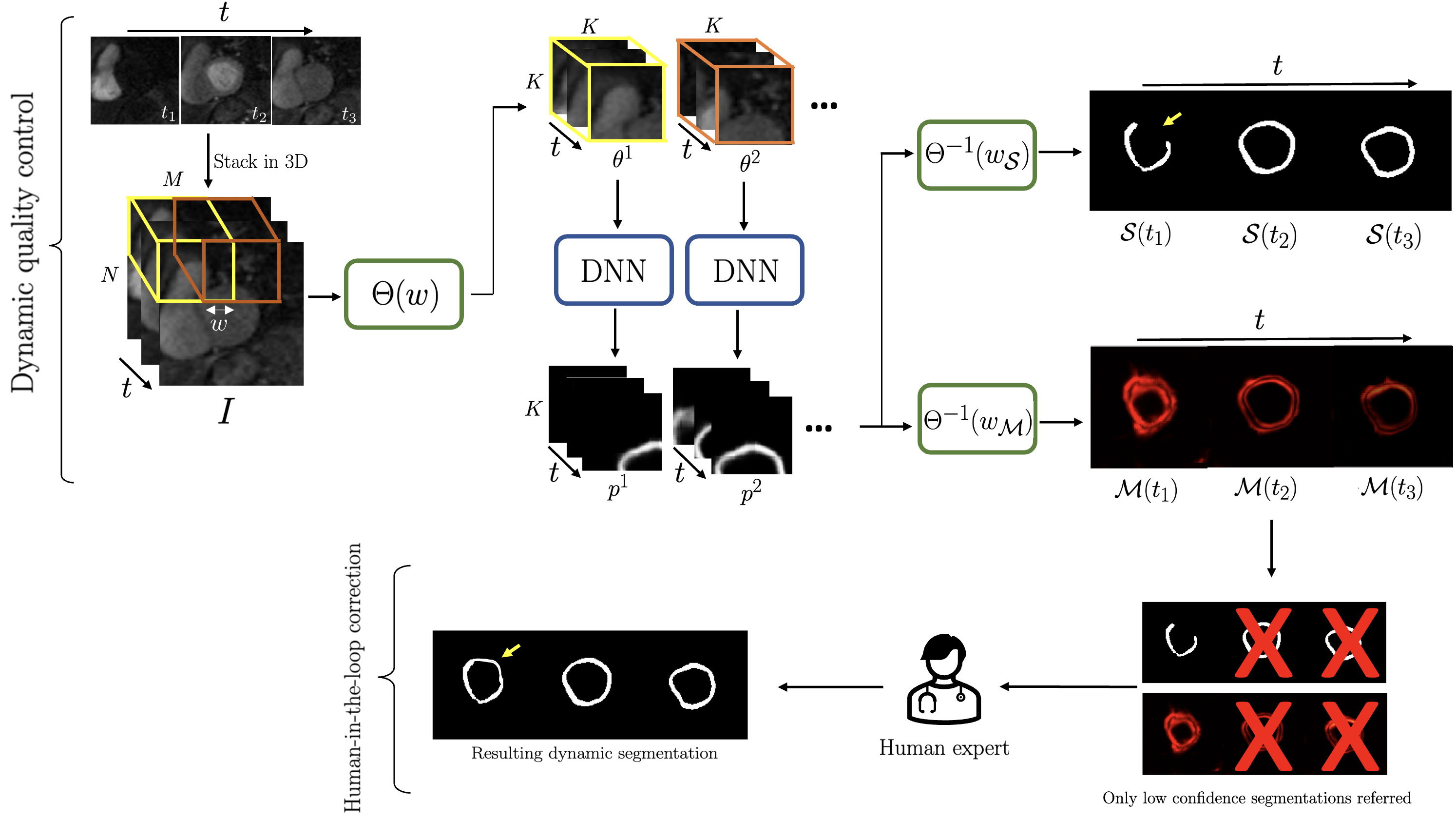}
\caption{Pipeline for the proposed dynamic quality control (dQC)-guided human-in-the-loop correction. With patch-based analysis, dQC map $\mathcal{M}$ is obtained and segmentation uncertainty is quantified as a normalized per-frame energy. Only low-confidence segmentations are referred to (and are corrected by) the human.}
\label{fig_method}
\end{figure}

\subsection{dQC-guided human-in-the-loop segmentation correction}\label{Sect:crit}
As shown in Fig. \ref{fig_method}, to demonstrate the utility of the proposed dQC metric, low confidence DNN segmentations in the test set, detected by the dQC metric $\mathcal{Q}^\mathrm{frame}$, were referred to a human expert for refinement who was instructed to correct two types of error: (i) anatomical infeasibility in the segmentation (e.g., non-contiguity of myocardium); (ii) inclusion of the right-ventricle, left-ventricular blood pool, or regions outside of the heart in the segmented myocardium. 

\subsection{DNN model training}\label{dnn_train}
We used a vanilla U-Net~\cite{unet} as the DNN time frames stacked in channels, and optimized cross-entropy loss using Adam. We used He initializer \cite{he}, batch size of 128, and linear learning rate drop every two epochs, with an initial learning rate of 5×10-4. Training stopped after a maximum of 15 epochs or if the myocardial Dice score of the validation set did not improve for five consecutive epochs. MATLAB R2020b (MathWorks) was used for implementation on a NVIDIA Titan RTX. CMRI images were preprocessed to a size of 128$\times$128$\times$25  after localization around the heart. Patch size of 64$\times$64$\times$25 was used for testing and training, with a patch combination stride of $w_\mathcal{S}$=16  and $w_\mathcal{M}$=2 pixels.

\section{Results}

\subsection{Baseline model performance}\label{init_res}

The ``baseline model'' performance, i.e., the DNN output without the human-in-the-loop corrections, yielded an average spatiotemporal (2D+time) Dice score of 0.767 ± 0.042 for the test set, and 16.2\% prevalence of non-contiguous segmentations, which is one of the criteria for failed segmentation (e.g., $\mathcal{S}(t_1)$ in Fig.~\ref{fig_method}) as described in Section \ref{Sect:crit}. Inference times on a modern workstation for segmentation of one acquired slice in the test set and for generation of the dQC-map were 3 seconds and 3 minutes, respectively. 

\subsection{Human-in-the-loop segmentation correction}
Two approaches were compared for human-in-the-loop framework: (i) referring the top 10\% most uncertain time frames detected by our proposed dQC tool (Fig. \ref{fig_method}), and (ii) randomly selecting 10\% of the time frames and referring them for human correction. The initial prevalence of non-contiguous (failed) segmentations among the dQC-selected vs. randomly-selected time frames was 46.8\% and 17.5\%, respectively. The mean 2D Dice score for dQC-selected frames was 0.607$\pm$0.217 and, after human expert corrections, it increased to 0.768$\pm$0.147 (p<$0.001$). On the other hand, the mean 2D Dice for randomly selected frames was initially 0.765$\pm$0.173 and, after expert corrections, there was only a small increase to 0.781$\pm$0.134 (p=n.s.). Overall, the human expert corrected 87.1\% of the dQC-selected and 40.3\% of the randomly-selected frames.

\begin{table}[h!]
\centering
\begin{tabular}{||c|c|c|c||} 
 \hline
 \textbf{} & \textbf{Baseline}  & \textbf{Random} & \textbf{~dQC-guided~} \\ [0.5ex] 
 \hline\hline
    Dice score & ~0.767 ± 0.042~ & ~0.768±0.042~ & ~0.781±0.039~ \\ 
    \hline
    ~failure prevalence~ & 16.2\% & 14.4\% & 11.3\% \\ [0.5ex] 
 \hline
\end{tabular}
\caption{Spatiotemporal (2D+time) cumulative results comparing the two methods for human-in-the-loop image segmentation.}
\label{table_3D}
\end{table}

Table \ref{table_3D} shows spatiotemporal (2D+time) cumulative results which contain all time frames including not selected frames for correction demonstrating that dQC-guided correction resulted in a notable reduction of failed segmentation prevalence from 16.2\% to 11.3\%, and in a significant improvement of the mean 2D+time Dice score. In contrast, the random selection of time frames for human-expert correction yielded a nearly unchanged performance compared to baseline. To calculate the prevalence of failed segmentations with random frame selection, a total of 100 Monte Carlo runs were carried out.

\subsection{Difficulty grading of DCE-CMRI time frames vs. $\mathcal{Q}^\mathrm{frame}$}
To assess the ability of the proposed dQC tool in identifying the ``most challenging'' time frames in a DCE-CMRI test dataset, a human expert (clinician) assigned ``difficulty grades'' to each time frame in our test set. The criterion for difficulty was inspired by clinicians' experience in delineating endo- and epicardial contours. Specifically, we assigned the following two difficulty grades: (i) Grade 1 (``high-grade difficulty''): both the endo- and epicardial contours are difficult to delineate from the surrounding tissue; (ii) Grade 0 (``moderate-to-low difficulty''): at most one of the endo- or epicardial contours are challenging to delineate.

\begin{figure}
\centering
\includegraphics[width=3in]{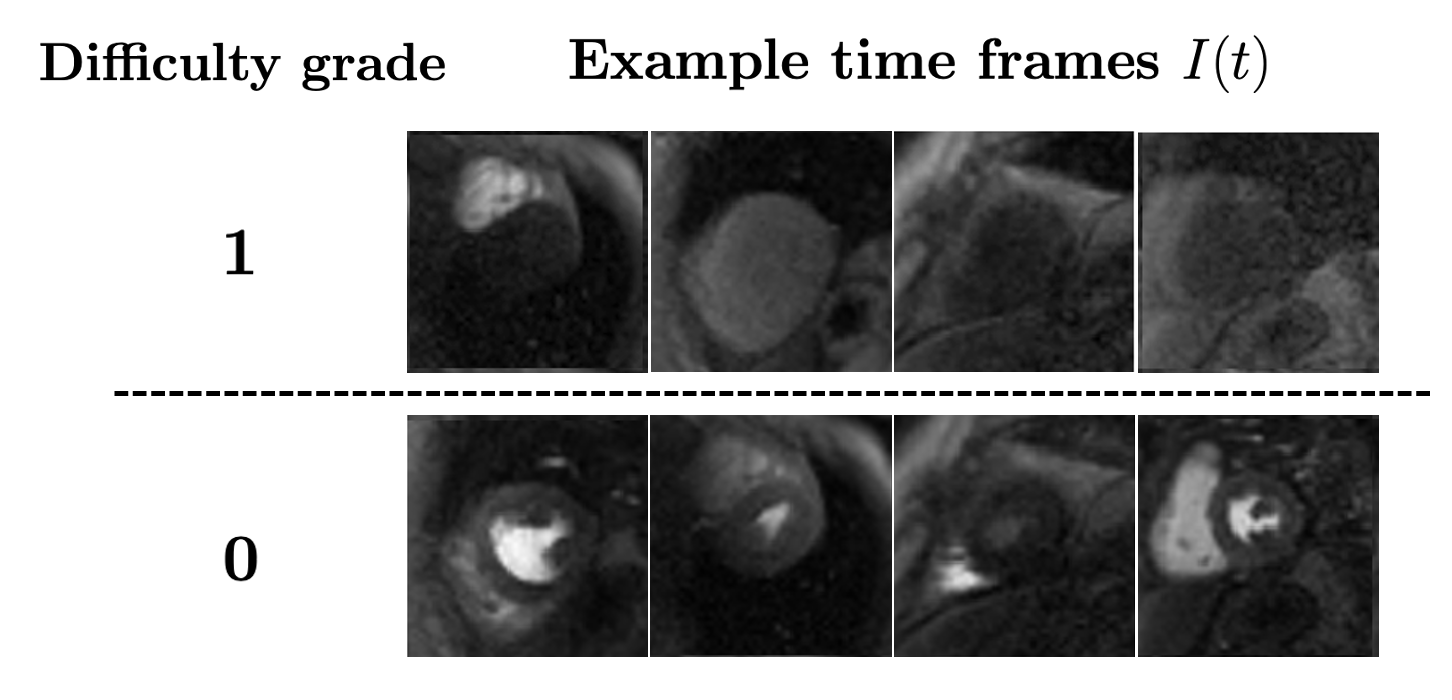}
\caption{Examples of DCE time frames corresponding to the two difficulty grades.} \label{fig_grades}
\end{figure}

To better illustrate, a set of example time frames from the test set and the corresponding grades are shown in Fig. \ref{fig_grades}. The frequency of Grade 1 and Grade 0 time frames in the test set was 14.7\% and 85.3\%, respectively. Next, we compared the agreement of $\mathcal{Q}^\mathrm{frame}$ values with difficulty grades through a binary classifier whose input is dynamic $\mathcal{Q}^\mathrm{frame}$ values for each acquired slice. Note that each $\mathcal{Q}^{\mathrm{frame}}$ yields a distinct classifier due to variation in heart size (hence in dQC maps) across the dataset. In other words, we obtained as many classifiers as the number of slices in the test set with a data-adaptive approach. The classifiers resulted in a mean area under the receiver-operating characteristics curve of 0.847$\pm$0.109, which indicates a strong relationship between dQC metric and the level of segmentation difficulty in a particular DCE-CMRI slice. 

\subsection{Representative cases}

Fig. \ref{rep_case1} shows two example test cases with segmentation result, dQC maps, and $\mathcal{Q}^{\mathrm{frame}}$. In (a), the highest $\mathcal{Q}^{\mathrm{frame}}$ was observed at $t$=22, coinciding with the failed segmentation result indicated by the yellow arrow (also see the peak in the adjoining plot). In (b), the segmentation errors in the first 6 time frames (yellow arrows) are accurately reflected by the $\mathcal{Q}^{\mathrm{frame}}$ metric (see adjoining plot) after which the dQC metric starts to drop. Around $t$=15 it increases again, which corresponds to the segmentation errors starting at $t$=16.

\section{Discussion and Conclusion}
In this work, we proposed a dynamic quality control (dQC) method for DNN-based  segmentation of dynamic (time resolved) contrast enhanced (DCE) cardiac MRI. Our dQC metric leverages patch-based analysis by analyzing the discrepancy in the DNN-derived segmentation of overlapping patches and enables automatic assessment of the segmentation quality for each DCE time frame. 

\begin{figure} 
\centering
\includegraphics[width=4in]{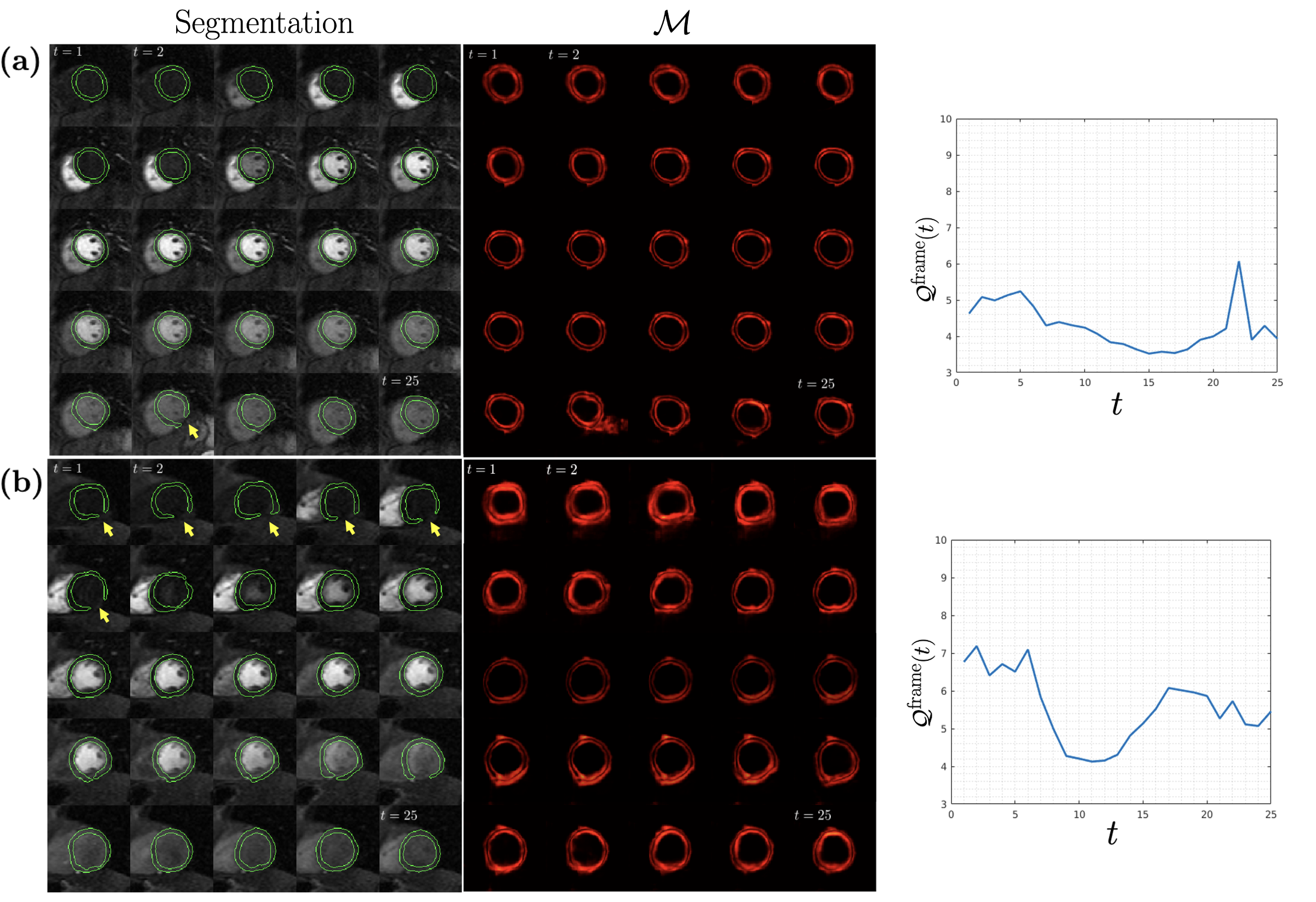}
\label{rep_case1}\caption{Two representative DCE-CMRI test cases are shown in along with segmentation, dQC maps $\mathcal{M}$, and the change of dQC metric $\mathcal{Q}^{\mathrm{frame}}(t)$ with time.}\label{rep_case1}
\end{figure}

To validate the proposed dQC tool and demonstrate its effectiveness in temporal localization of uncertain image segmentations in DCE datasets, we considered a human-A.I. collaboration framework with a limited time/effort budget (10\% of the total number of images), representing a practical clinical scenario for the eventual deployment of DNN-based methods in dynamic CMRI. 

Our results showed that, in this setting, the human expert correction of the dQC-detected uncertain segmentations results in a significant performance (Dice score) improvement. In contrast, a control experiment using the same number of randomly selected time frames for referral showed no significant increase in the Dice score, showing the ability of our proposed dQC tool in improving the efficiency of human-in-the-loop analysis of dynamic CMRI by localization of the time frames at which the segmentation has high uncertainty. In the same experiment, dQC-guided corrections resulted in a superior performance in terms of reducing failed segmentations, with a notably lower prevalence vs. random selection  (11.3\% vs. 14.4\%). This reduced prevalence is potentially impactful since quantitative analysis of DCE-CMRI data is sensitive to failed segmentations.

A limitation of our work is the subjective nature of the “difficulty grade” which was based on feedback from clinical experts. Since the data-analysis guidelines for DCE CMRI by the leading society \cite{scmrguide} do not specify an objective grading system, we were limited in our approach to direct clinical input. Any such grading system may introduce some level of subjectivity.

\subsubsection{Acknowledgement}
This work was supported by the NIH awards R01-HL153430 \& R01-HL148788, and the Lilly Endowment INCITE award (PI: B. Sharif).

%
%

\bibliographystyle{splncs04}
\bibliography{refs}

\end{document}